\documentstyle[manuscript,aps,epsfig,tighten,citesort]{revtex}

\newcommand{\tanbeta}    {${\rm tan\beta}$}
\newcommand{\beq}        {\begin{equation}}
\newcommand{\eeq}        {\end{equation}}
\newcommand{\beqa}       {\begin{eqnarray}}
\newcommand{\eeqa}       {\end{eqnarray}}
\newcommand{\hats}       {\hat{s}}
\newcommand{\hatmss}     {\hat{m_s}^2}
\newcommand{\hatmls}     {\hat{m_l}^2}
\begin{document}

\thispagestyle{empty}

\begin{flushright}
hep-ph/9902355
\end{flushright}
\vskip 1cm
%%%%%%%%%%%%%%%%%%%%%%%%%%%%%%%%%%%%%%%%%%%%%%%%%%%%%%%%%%%
%%%%%%%%%%%%    Title,authors & abstract
%%%%%%%%%%%%%%%%%%%%%%%%%%%%%%%%%%%%%%%%%%%%%%%%%%%%%%%%%%%

\begin{center}
{\Large \bf \boldmath
        $\tau$ Polarisation asymmetry in $B \to X_s~ \tau^+ \tau^-$ in
SUSY models with large $\tan\beta$ }

\vskip 2cm
{ \bf
  S. Rai Choudhury \footnote{Email : src@ducos.ernet.in},
  Naveen Gaur \footnote{Email : naveen@physics.du.ac.in,~
ngaur@ducos.ernet.in}, 
  Abhinav Gupta \footnote{Email : abh@ducos.ernet.in} } \\
{\em  Department of Physics \& Astrophysics,\\
  University of Delhi, \\
  Delhi 110 007, \\
  India.  \\ }
\end{center}
\vskip 1cm
\begin{center}
{\large Abstract}
\end{center}
%%%%%%%%%%%%%%%%%%%%%%%%%%%%%%%%%%%%%%%%%%%%%%%%%%%%%%%%%%%
%%%%%%%%%%%%%            Abstract
%%%%%%%%%%%%%%%%%%%%%%%%%%%%%%%%%%%%%%%%%%%%%%%%%%%%%%%%%%%
  Rare B decays provides an opportunity to probe for new physics beyond
the standard model. the effective Hamiltonian for the decay $b \to s l^+
 l^-$ predicts the characteristic polarization for the final state
lepton. Lepton 
polarization has, in addition to a longitudinal component $P_L$ , two
orthogonal components $P_T$ and $P_N$ lying in and perpendicular to
the decay plane. In this article we perform a study of the
$\tau$-polarisation asymmetry in the case of SUSY models with large
$\tan\beta$ in the inclusive decay $B \to X_s \tau^+ \tau^-$
%\end{abstract}

\vfill
\pagebreak

%%%%%%%%%%%%%%%%%%%%%%%%%%%%%%%%%%%%%%%%%%%%%%%%%%%%%%%%%%%
%%%%%%%%%%%%      Main file of paper
%%%%%%%%%%%%%%%%%%%%%%%%%%%%%%%%%%%%%%%%%%%%%%%%%%%%%%%%%%%
	Recent progress in experiment and theory has made flavor
changing neutral current (FCNC) B decays a stringent test of the standard
model (SM) and a powerful probe of physics beyond standard model.
The first observations \cite{cleo1} of the inclusive and exclusive 
radiative decays $B \to X_s \gamma$ and $B \to K^* \gamma$ have placed 
the study of rare B decays on a new footing.
The observation of $b \to s \gamma$ by CLEO puts very strong
constraints on various new physics beyond standard model.In the case of $B
\to X_s \gamma$ CLEO observation gives very strong constraints on
charged Higgs boson mass in the two Higgs doublet model.But in the minimal
supersymmetric standard model (MSSM) these constraints becomes a bit 
relaxed because of various cancellations between different superparticle
contributions. It is therefore important to study the sensitivity of other
FCNC processes to SUSY.
\par Recently the inclusive decay of $B \to X_s l^+ l^-$
\cite{cho1,ali1} received 
considerable attention as a testing ground of SM and new physics.
The experimental situations of these decays is
very promising with $e^+ e^-$ and hadronic colliders closing on the
observation of exclusive models with $l = \mu$ and $e$ final states
respectively. In this decay we can observe various kinematical
distributions associated with a final state lepton pair such as lepton
pair invariant mass spectrum, lepton pair forward backward asymmetry
etc. Recently another observable, $\tau$ polarisation asymmetry, for
the $B \to X_s \tau^+ \tau^-$ mode has also been proposed by Hewett
\cite{hewett1} which can again be used for more strict checking of
effective Hamiltonian governing the decay.
In another work \cite{kruger1} attention has been drawn to 
the fact that apart
from Longitudinal polarization of lepton there can be two other
orthogonal components of polarisations which are proportional to
$m_l/m_b$ and hence are important for $\tau$. These components
of polarizations namely the component in the decay plane ($P_T$, transverse
polarization) and the component normal to decay plane ($P_N$, normal
polarization).
\footnote{different combinations of the Wilson coefficients
describing the decay and are thus useful for comparing theory with
experimental data.}
In this paper we will try to examine the sensitivity of
these observables with respect to new physics i.e MSSM.
%%%%%%%%%%%%%%%%%%%%%%%%%%%%%%%%%%%%%%%%%%%%%%%%%%%%%%%%%%%%%%%
%%%%%%%%%%%%%%  Modification 18-05-1999  %%%%%%%%%%%%%%%%%%%%%%
%%%%%%%%%%%%%%%%%%%%%%%%%%%%%%%%%%%%%%%%%%%%%%%%%%%%%%%%%%%%%%%
\par Among models for physics beyond standard model supersymmetry
(SUSY) is considered to be the most promising candidate. The minimal
extension of the standard model (MSSM) involves chiral superfields
$Q,U^c,D^c,L,E^c,H_1$ and $H_2$ which transforms under $SU(3)_c
\times SU(2)_L \times U(1)_Y$ as 
\beqa
\label{newone}
Q \equiv (3,2,1/2),\quad &&\quad  U^c \equiv (\bar{3},1,-2/3) 
\nonumber \\
D^c \equiv (\bar{3},1,1/3),\quad  && \quad  L \equiv (1,2,-1/2)    
\nonumber \\ 
E^c \equiv (1,1,1),\quad && \quad H_1 \equiv (1,2,-1/2) \nonumber \\
&H_2 \equiv (1,2,1/2)& \nonumber \\
\eeqa
The superpotential in MSSM in terms of these superfields are 
\beq
\label{newtwo}
W = h_U^{ij} Q_i U_j^c H_2 + h_D^{ij} Q_i D_j^c H_1 + \mu H_1 H_2 +
h_e^{ij} L_i E_j^c H_1
\eeq
 where {\it i,j} denote generation indices ($i,j = 1,2,3$), and $\mu$
and h's are 
parameters of MSSM. Supersymmetry is broken softly in MSSM. At a large 
grand-unified scale $M_G$ the bilinear terms have the structure
\beq
\label{newthree}
M_{\rm soft}^{(2)} = \sum_i m_i^2 |y_i|^2 + {1 \over 2} \sum_j ( M_j
\lambda_i \lambda_j + {\rm h.c.})
\eeq
 where ${y_i} 's$ are the scalar components of the chiral superfields and 
$\lambda_1,\lambda_2,\lambda_3$ are the two component gaugino fields
of ${\rm U(1)_Y,SU(2)_L}$ and ${\rm SU(3)_c}$, $m_i,M_i's$ are parameters.
The trilinear soft breaking term is 
\beq
\label{newfour}
M_{\rm soft}^{(3)} = m A [ h_U Q^s U^s H_2^s + h_D Q^s D^s H_1^s + h_E L^s 
E^s H_1^s + B m \mu H_1^s H_2^s + {\rm h.c.}]
\eeq
where the superscript s denote the scalar component of the
corresponding superfield and the generation index is suppressed in
Eq.(\ref{newfour}). A and B are constants and m is a scale factor. At scale 
$\sim M_W$, the ${\rm SU(2)_L \times U(1)_Y}$ symmetry is broken
spontaneously by the $H_1,H_2$ developing a nonzero vacuum
expectation value.
\beq
\label{newfive}
\langle H_1 \rangle = \left( \begin{array}{c}
                      v_1 \\
                      0   \end{array} \right) ~~~ , ~~~
\langle H_2 \rangle = \left( \begin{array}{c}
                      0 \\
                      v_2   \end{array} \right)
\eeq
The quantity ${\rm tan\beta} = v_2/v_1$, thus enters as another parameter in 
MSSM.
\par The MSSM in its general form has far too many parameters for it
to be used in phenomenology in any meaningful way. Most applications
have considered MSSM in the context of minimal spontaneously broken $N 
= 1$ supergravity (SUGRA). This implies that at the Planck scale all
the scalar masses have an universal value ($m_i = m$) as do the
gauginos ($M_i = M$). At $M_G$ we thus have five parameters (apart from 
gauge and Yukawa couplings and \tanbeta) $A, B, \mu, m ~{\rm and} M$. Using
renormalization group equations these parameters can be scaled down to 
the scale $M_W$. The condition that at scale $M_W$, the ${\rm SU(2)_L \times
U(1)_Y}$ symmetry breaks down to ${\rm U(1)_{em}}$, via the spontaneous
symmetry breaking (SSB) condition
Eq.(\ref{newfive}), reduces the number of independent parameters
2. However, as discussed in \cite{goto1}, we use a more relaxed SUGRA 
model which requires the degeneracy of soft SUSY-breaking mass in the
scalar squark sector and separately in the higgs boson sector thus in
Eq. (\ref{newthree}) $m_i = m_0$ for squarks and $m_i = \Delta_0$ for
the higgs boson. Thus as has been discussed by Goto 
{\em et al.} \cite{goto1}, is sufficient to ensure an important constraint,
namely adequate suppression of $K^0 - \bar{K^0}$ mixing.
\par Ths MSSM has been used to study various rare decays such as  $b \to s
l^+ l^-, b \to s \nu \bar{\nu}, K^0 \to \pi^0 l^+ l^-$ using the known 
results of $b \to s \gamma$ \cite{cleo1} as a constraint on the
parameter space \cite{lopez1,src1}. It was also observed that 
very large value of \tanbeta ~is still allowed \cite{bert1,lopez1,src1}
It has been pointed out recently by
\cite{dai1,src1} that for large \tanbeta, which is allowed by the
constraining condition, the process $b \to s l^+ l^-$ can also proceed 
via exchange of neutral Higgs bosons (NHB) $h^0,H^0$ and $A^0$. These
exchanges lead to additional amplitudes which scale such as $m_b m_l 
tan^3\beta$ and this can give considerable enhancement of processes
like $B_s \to \mu^+ \mu^-, B \to X_s l^+ l^-$, 
etc. \cite{dai1,src1}. For $l = \tau$, these NHB contributions for
large \tanbeta will be even more significant. In this paper we will try to 
estimate $\tau$-polarization parameters including NHBs contributions.
\par We start by writing down the QCD improved effective hamiltonian
for the process $B \to X_s l^+ l^-$  \cite{dai1}:
\beqa
\label{one}
{\cal H} 
&=& 
\frac{\alpha G_F}{\sqrt{2} \pi} V_{tb}V_{ts}^*
 \Bigg[ C_9^{eff} (\bar{s} \gamma_\mu P_L b) \bar{l}\gamma^\mu l
     + C_{10}(\bar{s} \gamma_\mu P_L b) \bar{l} \gamma^\mu \gamma^5 l
     - 2 C_7^{eff} \bar{s} i \sigma_{\mu \nu} \frac{p^\nu}{p^2}
       (m_b P_R + m_s P_L)b \bar{l} \gamma^\mu l \nonumber \\
&& + C_{Q_1} (\bar{s} P_R b)\bar{l} l 
   + C_{Q_2} (\bar{s}P_R b)\bar{l} \gamma_5 l
\Bigg]
\eeqa
with $P_{L,R} = {1 \over 2}(1 \mp \gamma_5), p = p_+ + p_-$ sum of the 
momentum of $l^+$ and $l^-$,$C_9^{\rm eff},~C_{10}$ and $C_7^{\rm eff}$ are
Wilson coefficients given in \cite{buras1,kruger1}. $C_{Q_1}$ and
$C_{Q_2}$ are new Wilson coefficients which are absent in the standard 
model but arises in MSSM due to NHB exchange. Their values are given
in Ref.\cite{dai1}.The C's all receive contributions from diagrams
involving SUSY particles. However as has been pointed out in
refs. \cite{lopez1,goto1}, the various SUSY contributions to $C_7,C_9$ and
$C_{10}$ have large cancellations amongst themselves leading to only
mild changes in their values relative to SM. $C_{Q_1}$ and $C_{Q_2}$'s 
,which have only SUSY contributions, for certain regions of allowed
parameter space (space allowed by $b \to s \gamma$) can be comparable
to magnitude of $C_{10}$. We however, include the SUSY contributions
to all Wilson coefficients as given in Ref.\cite{lopez1,goto1} for our
numerical estimates. 
\par $B \to X_s l^+ l^-$ also receives large long distance
contributions from tree level process associated with $c\bar{c}$ 
resonances in intermediate
states i.e. with chain reaction $B \to X_s + \Psi \to
X_s l^+ l^-$. These resonant contributions  can be incorporated into 
lepton pair invariant mass spectrum according to prescription of
Ref.\cite{deshpande1} by employing Breit-Wigner form of the resonance
propogator. This produces additional contribution to $C_9^{\rm eff}$ of
the form 
\beq
\label{two}
\frac{-3 \pi}{\alpha^2 } 
\sum_{V = J/\psi,\psi',..}
\frac{M_V Br(V \to l^+ l^-)\Gamma_{\rm total}^V}{(s - M_V^2) 
+ i \Gamma_{\rm total}^V M_V}
\eeq
where the properties of the vector mesons are given in a table in
Ref.\cite{kruger1}.
There are six known resonances in the $c\bar{c}$ system that can
contribute to the decay modes $B \to X_s l^+ l^-$. Of these, all
except the lowest $J/\psi(3097)$ contribute to the channel $B \to X_s
\tau^+ \tau^-$, for which the invariant mass of lepton pair is $s >
4 m_\tau^2$, i.e. greater then $\tau$ pair production threshold
\footnote{as given in references the prescription eqn(\ref{two}) for the
resonance contribution implies an inclusive direct $J/\psi$ production
rate $Br(B \to J/\psi X_s) = 0.15 $ that is $\sim 5$ times smaller 
than the measured $J/\psi$ rate.This is corrected by the introduction 
of a phenomenological factor of $\kappa_v \approx 2$ multiplying the
Breit-Wigner function in (\ref{two}).For our results we use $\kappa_v
= 2.35$ }.
\par The differential decay rate for $B \to X_s \tau^+ \tau^-$
is then 
\beq
\label{three}
\frac{dB(B \to X_s \tau^+ \tau^-)}{d\hat{s}}
= \frac{G_F^2 m_b^5}{192 \pi^3} \frac{\alpha^2}{4 \pi^2} |V_{tb}
V_{ts}^*|^2 \lambda^{1/2}(1,\hat{s},\hat{m_s}^2) 
\sqrt{1 - \frac{4 \hat m_l ^2}{\hat{s}}} \bigtriangleup
\eeq
where factors $\lambda$ and $\bigtriangleup$ are defined by
\beq
\label{four}
\lambda(a,b,c) = a^2 + b^2 + c^2 - 2(a b + b c + a c)
\eeq
and
\beqa
\label{five}
\bigtriangleup 
&=& 
\left\{
 \left(  {4 \over \hat{s}}|C_7^{\rm eff}|^2 F_1(\hat{s},\hat{m_s}^2)
   + 12 Re(C_7^{\rm eff} C_9^{\rm eff}) F_2(\hat{s},\hat{m_s}^2)
 \right)(1 + \frac{2 \hatmls}{\hats})   \right.        \nonumber  \\
&& \left. + |C_9^{\rm eff}|^2 F_3(\hats,\hatmss,\hatmls) 
    + |C_{10}|^2 F_4(\hat{s},\hat{m_s}^2,\hat{m_l}^2) \right. \nonumber \\
&& \left. +{3 \over 2}|C_{Q_1}|^2 F_5(\hats,\hatmss)(\hats - 4 \hatmls)
  + {3\over 2}|C_{Q_2}|^2 F_6(\hats,\hatmss)   \right.    \nonumber \\
&& \left. + 6 C_{Q_2} C_{10}  \hat{m_l} F_7(\hats,\hatmss) \right\}
\eeqa
with
\beqa
\label{six}
F_1(\hats,\hatmss) &=& 2 (1 + \hatmss)(1 - \hatmss)^2 
       - \hats (1 + 14 \hatmss + \hat{m_s}^4)  
       - \hat{s}^2(1 + \hatmss) \\
F_2(\hats,\hatmss) &=& (1 - \hatmss)^2 - \hats ( 1 + \hatmss) \\
F_3(\hats,\hatmss,\hatmls) 
  &=&  (1 - \hatmss)^2 + \hats (1 - \hatmss) - 2 \hats^2 
   + \frac{2 \hatmls}{\hats}( (1 + \hatmss) \hats + (1 -
       \hatmss)^2 - 2 \hats)                                   \\
F_4(\hats,\hatmss,\hatmls) 
  &=&  (1 - \hatmss)^2 + \hats (1 - \hatmss) - 2 \hats^2 
   + \frac{2 \hatmls}{\hats}(- 5 (1 + \hatmss) \hats + (1 -
       \hatmss)^2 + 4 \hats)                                      \\
F_5(\hats,\hatmss) 
  &=& 1 + \hatmss - \hats                                     \\
F_6(\hats,\hatmss) 
  &=& \hats (1 + \hatmss - \hats)                 \\
F_7(\hats,\hatmss) 
  &=& 1 - \hatmss - \hats                         
\eeqa
where we have used the notion that $\hats = p^2/m_b^2$,$\hat{m_i} =
m_i/m_b$. This matches with the result of Ref.\cite{dai1}.
\par	Now we discuss the final state lepton polarization.
The polarized crosssections are obtained by introducing the spin
projection operator. For $l^-$,
\beq
P_\mp = {1\over2}(1 + \gamma_5 \not\!N_i) ~~  ; ~~ i = L,T,N
\eeq
$(N_\mu)_i$ here are four-vectors satisfy $N p_- = 0$ and $N^2 = -1$. In
general,
\beqa
(N_\mu)_L &=& \left( \frac{|\hat{p}_-|}{m_l}
            ,\frac{p_-^0}{m_l}\hat{e}_L \right),  \\
(N_\mu)_T &=& (0,\hat{e}_T), \\
(N_\mu)_N &=& (0,\hat{e}_N)
\eeqa
where
\beqa
\hat{e}_L &=& \hat{p}_-,      \\
\hat{e}_N &=& \frac{ \vec{p}_s \times \vec{p}_-}
{|\vec{p}_s \times \vec{p}_-|},       \\
\hat{e}_T &=& \hat{e}_N \times \hat{e}_L
\eeqa
with $\vec{p}_-$ and $\vec{p}_s$ being the three-momentum of $l^-$ and
s-quark in CM frame of $l^+ l^-$.
\par The differential decay rate of $B \to X_s \l^+ l^-$ for any given 
spin direction $\hat{n}$ of lepton $l^-$ may then be written as:
\beq
\label{eight}
\frac{dB(n)}{d\hats} = {1\over2}
 \left(\frac{dB}{d\hats}\right)_{\rm unpol}
  \Bigg[ 1 + (P_L \hat{e}_L + P_T \hat{e}_T + \hat{e}_N) \hat{n} \Bigg],
\eeq
where $P_L,P_T$ and $P_N$ are functions of $\hats$ which gives
longitudinal, transverse and normal polarization components of
polarization respectively.The Polarization component $P_i(i=L,T,N)$ is
obtained by evaluating :
\beq
\label{nine}
P_i(\hats) = \frac{dB(\hat{n} = \hat{e}_i)/d\hats - 
                    dB(\hat{n} = - \hat{e}_i)/d\hats }
                  {dB(\hat{n} = \hat{e}_i)/d\hats +
                    dB(\hat{n} = - \hat{e}_i)/d\hats }
\eeq
The results obtained using the effective hamiltonian (\ref{one}) is :
\beqa
\label{ten}
P_L(\hats) 
 &=& \sqrt{1 - \frac{4\hatmls}{\hats}}
 \Bigg[     12 C_7^{\rm eff} C_{10} ( (1 - \hatmss)^2 
        - \hats(1 + \hatmss))          \nonumber \\
 && + 2 Re(C_9^{\rm eff} C_{10}) ( (1 - \hatmss)^2 + \hats(1 + \hatmss)
                               - 2 \hats^2)    \nonumber \\
 && + 6 C_{Q_1}C_{10}\hat{m}_l ( - 1 + \hatmss + \hats) 
   + 3 C_{Q_1}C_{Q_2}( -1 - \hatmss + \hats)\hats
\Bigg]/\bigtriangleup                                        \\
%%%%%%%%%%%%%%%%%%%%%%%%%%
P_T(\hats) 
 &=& \frac{3 \pi \hat{m}_l}{2 \sqrt{\hats}} \lambda^{1/2}(1,\hats,\hatmss)
 \Bigg[  2 C_7^{\rm eff} C_{10}(1 - \hatmss)      
        - 4 Re(C_7^{\rm eff} C_9^{\rm eff})(1 + \hatmss)  \nonumber \\
 && - {4\over \hats}|C_7^{\rm eff}|^2(1 - \hatmss)^2  
   + Re(C_9^{\rm eff} C_{10})(1 - \hatmss) - |C_9^{\rm eff}|^2 \hats  \nonumber \\
 &&  - {1\over2}C_{Q_1}C_{10} \frac{4 \hatmls - \hats}{\hat{m}_l}
     + C_{Q_2}C_7^{\rm eff} \frac{\hats}{\hat{m}_l}    
   + \frac{1}{2} Re(C_{Q_2}C_9^{\rm eff}) \frac{\hats}{\hat{m}_l}
 \Bigg]/\bigtriangleup                    \\
%%%%%%%%%%%%%%%%%%%%%%%%%%
P_N(\hats)
 &=& \frac{3 \pi \hat{m}_l}{2 \bigtriangleup}\sqrt{\hats}
     \lambda^{1/2}(1,\hats,\hatmss)\sqrt{1 - \frac{4\hatmls}{\hats}}
      Im({C_9^{\rm eff}}^*)({1\over2}C_{Q_1} + C_{10}\hat{m}_l)
\eeqa
Expressions of $P_L$ and $P_T$ matches with Ref.\cite{kruger1} if
$C_{Q_1}$ and $C_{Q_2}$ are absent, i.e. no NHB exchange effects. But
$P_T$ disagrees with Ref.\cite{kruger1} for a factor of 2 multiplying 
in term $C_7^* C_{10}$. 
Let us now focus our attention on the parameter space. Apart from gauge 
and Yukawa couplings, we have in the ``relaxed'' SUGRA model discussed
above, six parameters  $m_0,M,\Delta_0,A,B$ and $\mu$ at Planck
scale. Use of renormalization group equations(RGE) allows one to
evolve these parameters down to the electroweak scale $M_W$. At that
scale the ${\rm SU(2)_L \times U(1)_Y}$ spontaneously breaks down to
${\rm U(1)_{em}}$ Eq.(\ref{newfive}) $v_1$,$v_2$ determined in the tree
approximation by the Higgs boson 
potential with all its parameters scaled down to $M_W$. $M_Z$ is
related to $v_1$ and $v_2$ by 
\beq
M_Z^2 = {1 \over 2} (g^2 + g'^2) (v_1^2 + v_2^2),
\eeq
with $g, g'$ being respectively the ${\rm SU(2)_L ~and ~U(1)_Y}$ gauge
couplings. Thus, for a given value of $tan\beta = v_2/v_1$, and with
all SM parameters given we have effectively four free parameters,
which will be further subject to constraints arising out of the known
limits on $b \to s \gamma$.
\par
\begin{figure}
\epsfig{file=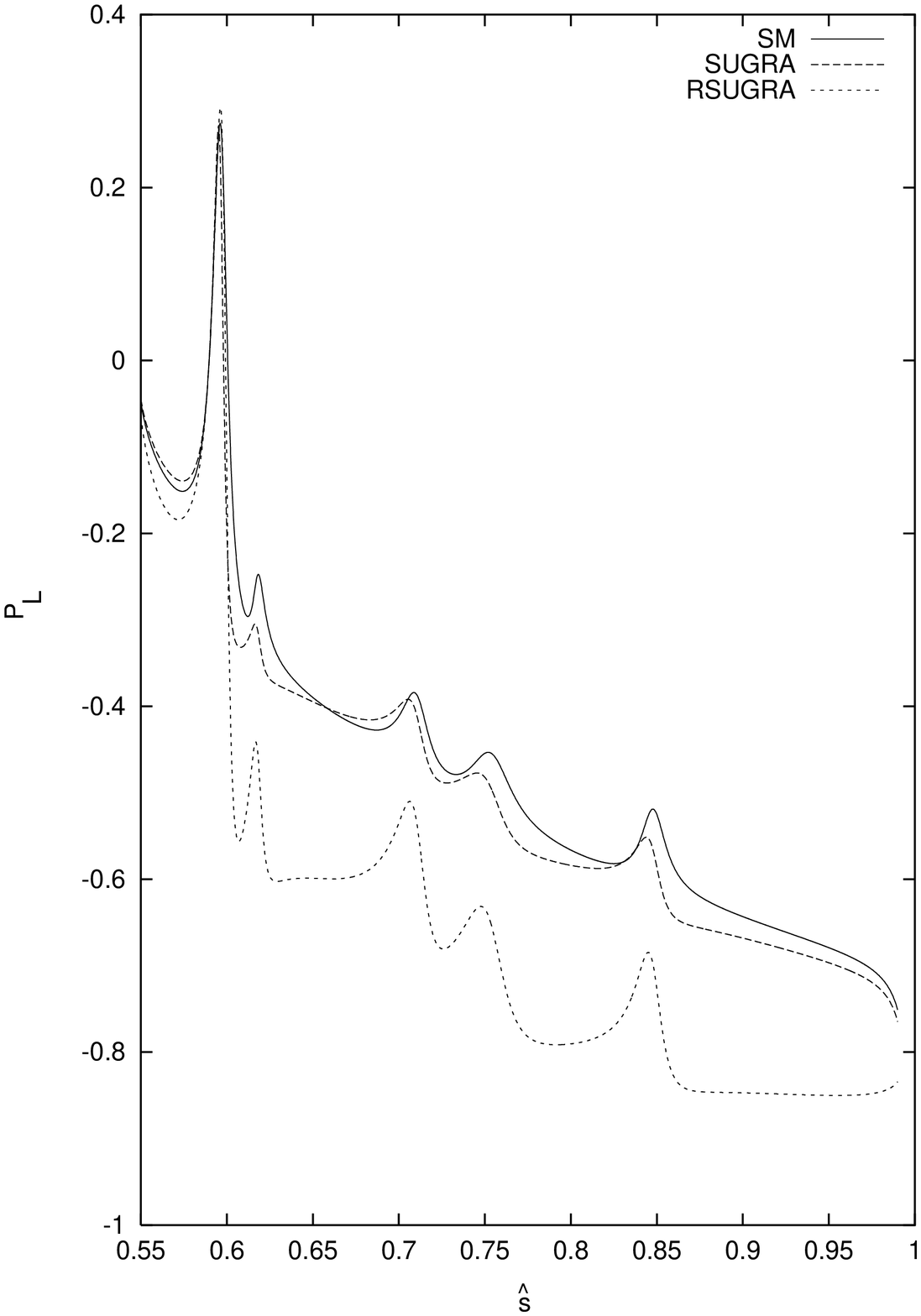,width=6in,height=7in}
\vskip 1cm
\caption{Longitudinal Polarisation asymmetry with $\hat{s}$
, parameters taken are  $tan\beta = 30 $ ; $ m = M = 130 ~;~ A = - 1$.
 For relaxed 
SUGRA (RSUGRA) model $m_A= 120$. All masses are in GeV}
\label{figure1}
\end{figure}

\begin{figure}
\epsfig{file=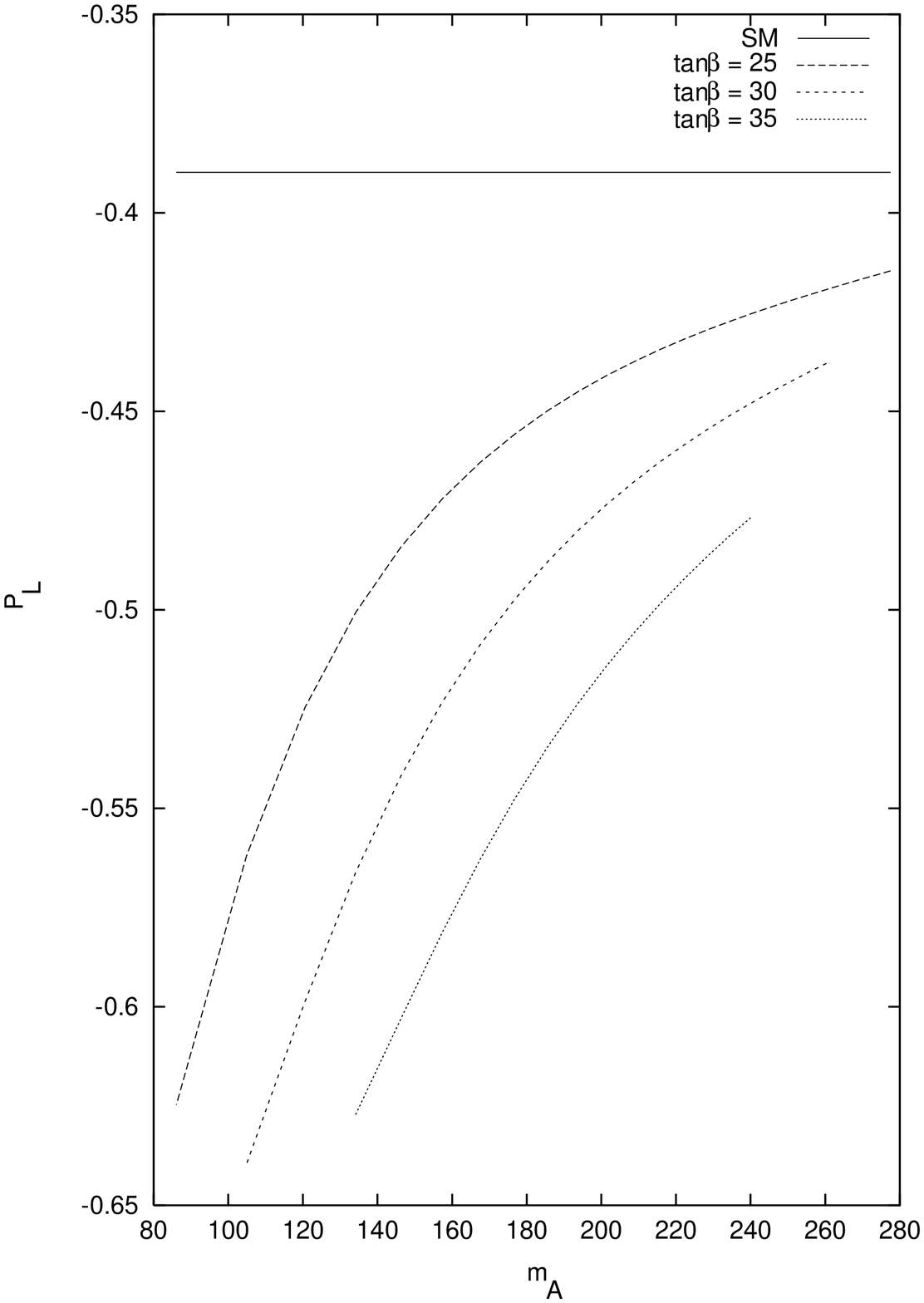,width=6in,height=7in}
\vskip 1cm
\caption{Longitudinal Polarisation asymmetry with $m_A$ in relaxed
SUGRA model.Other parameters are : $m = M = 130 ; A = - 1 ;
\hat{s} = 0.65$.All masses are in GeV}
\label{figure2}
\end{figure}

\begin{figure}
\epsfig{file=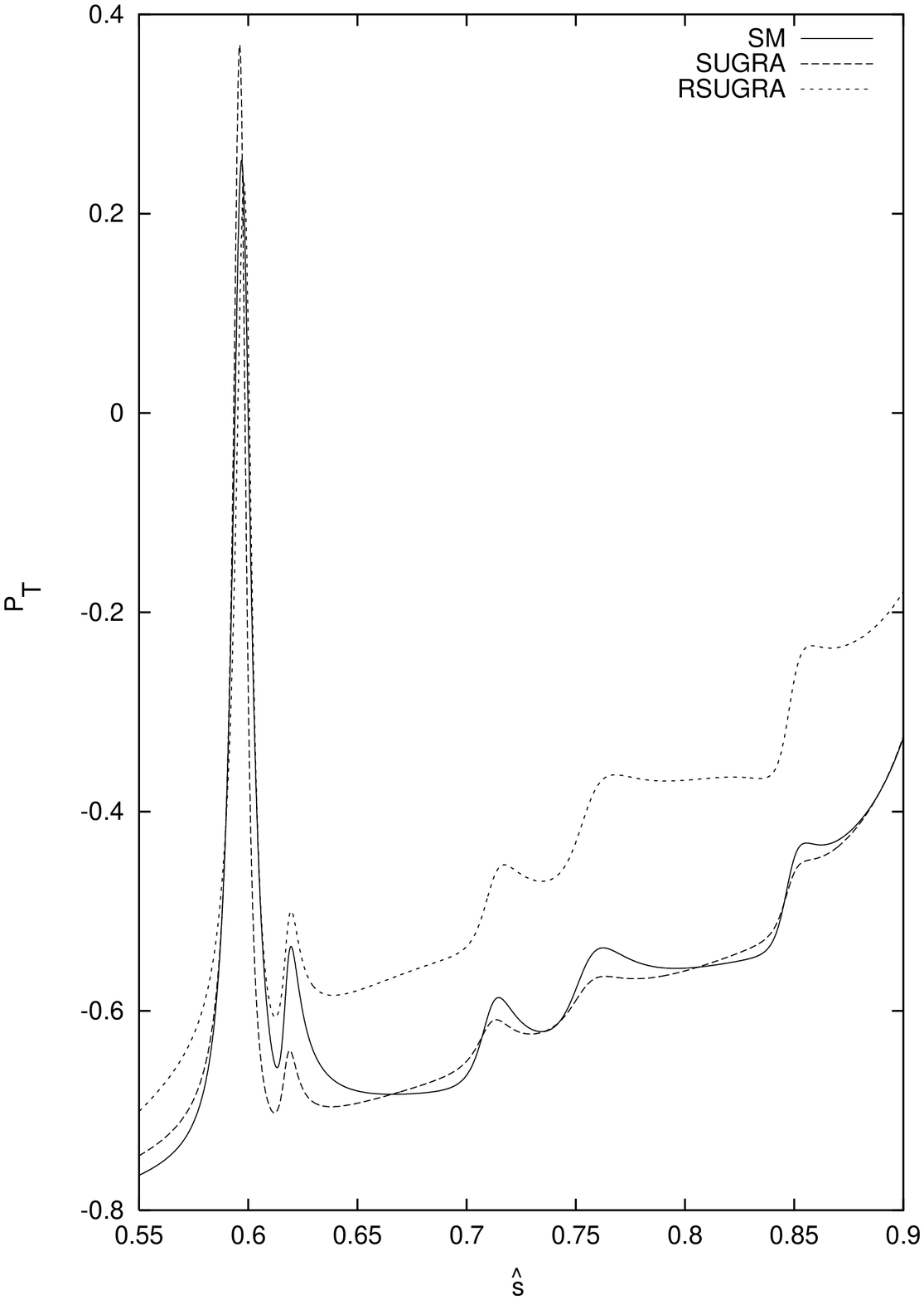,width=6in,height=7in}
\vskip 1cm
\caption{Transverse Polarisation asymmetry with $\hat{s}$. Other
parameters are : $tan\beta = 30 $ ; $m = M = 130 ~;~ A = - 1$. 
For relaxed SUGRA (RSUGRA)
model $m_A = 120$.All masses are in GeV.}
\label{figure3}
\end{figure}

\begin{figure}
\epsfig{file=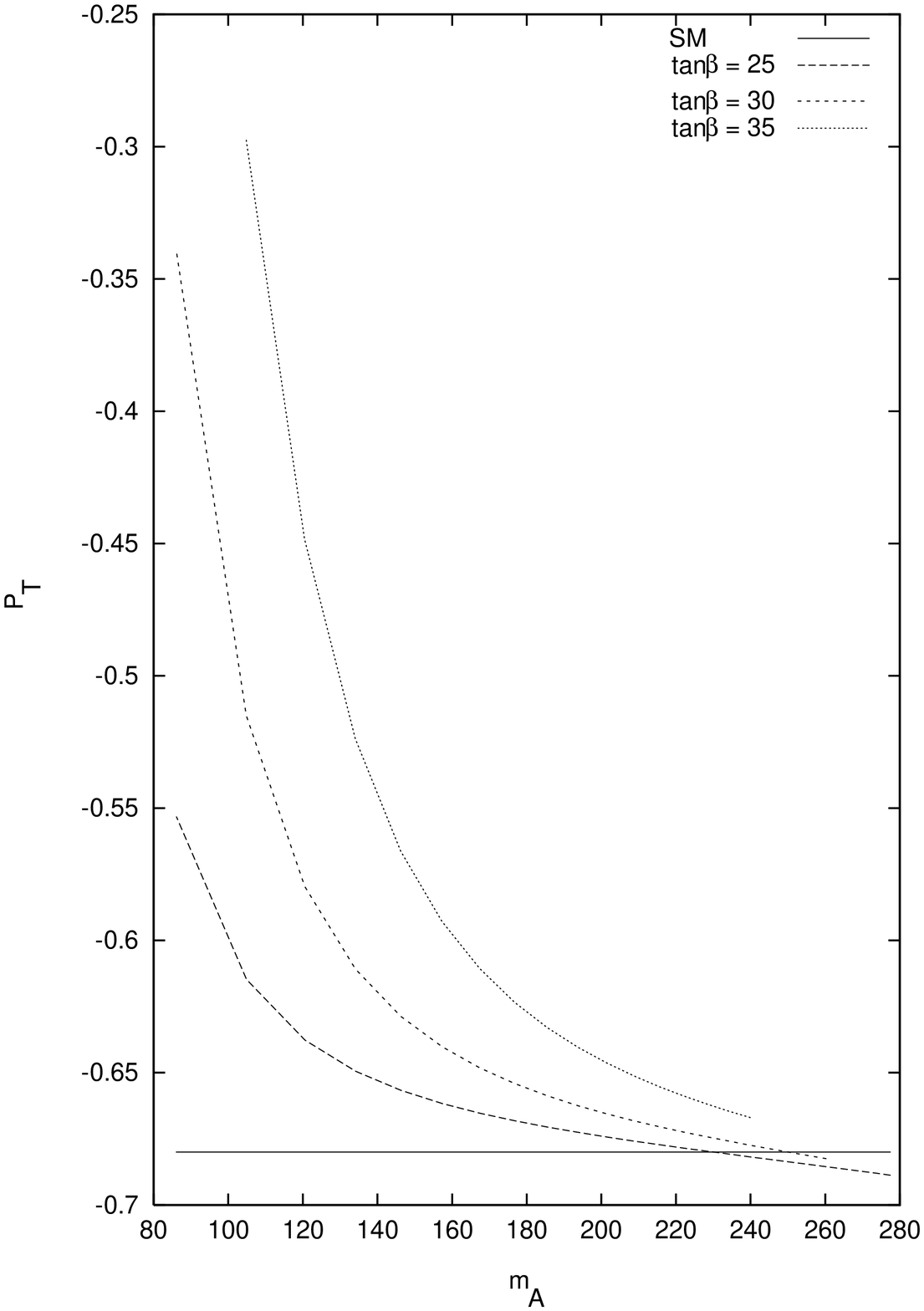,width=6in,height=7in}
\vskip 1cm
\caption{Transverse Polarisation asymmetry with $m_A$ in relaxed SUGRA
model.Other parameters are : $m = M = 130; A = - 1 ; \hat{s} =
0.65$. All masses are in GeV}
\label{figure4}
\end{figure}

\begin{figure}
\epsfig{file=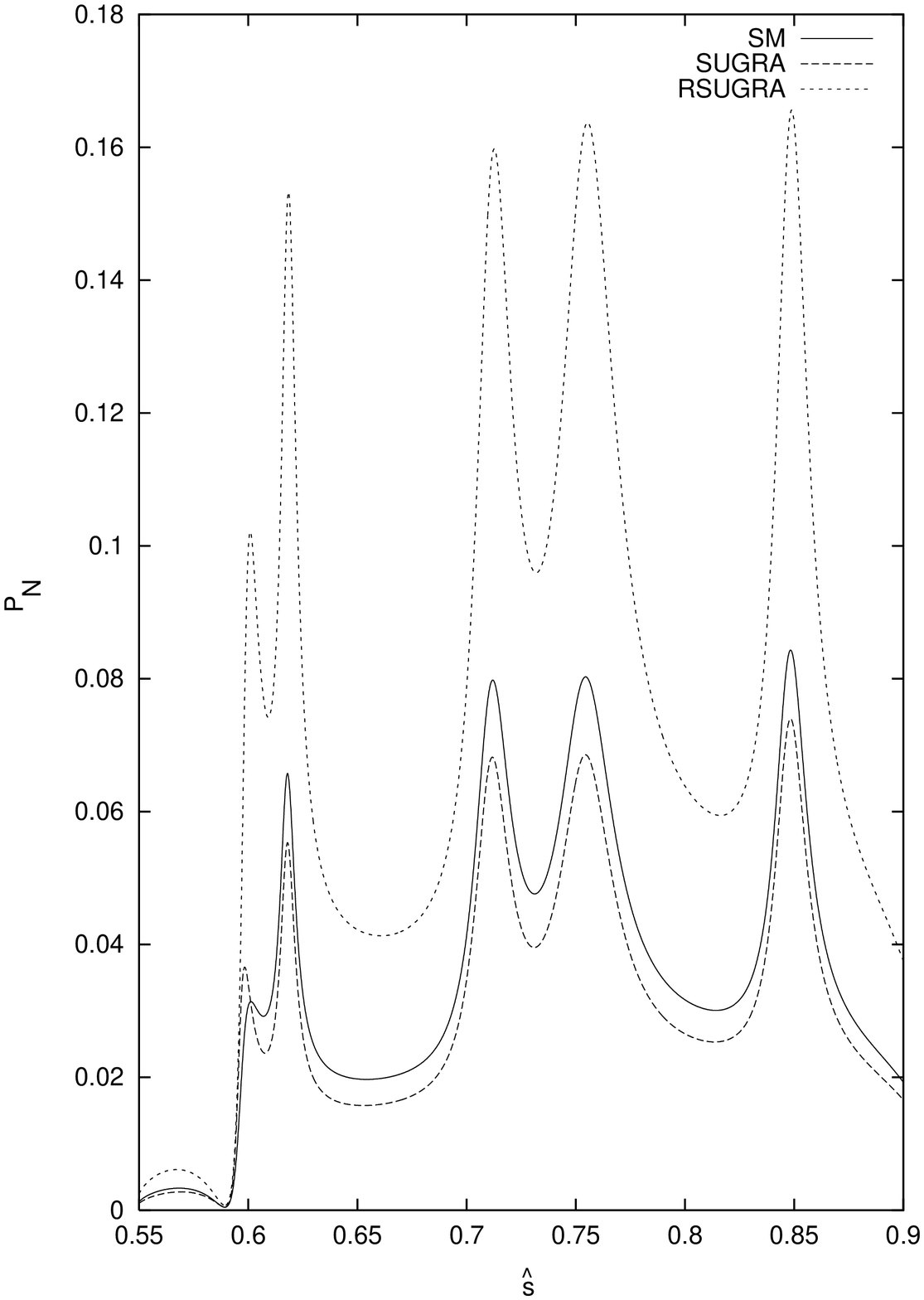,width=6in,height=7in}
\vskip 1cm
\caption{Normal Polarisation asymmetry with $\hat{s}$. Other
 parameters are : $ tan\beta = 30$  ; $ m = M = 130 ; A = - 1$ 
For relaxed SUGRA (RSUGRA)
model $ m_A = 120 $. All masses are in GeV.}
\label{figure5}
\end{figure}

\begin{figure}
\epsfig{file=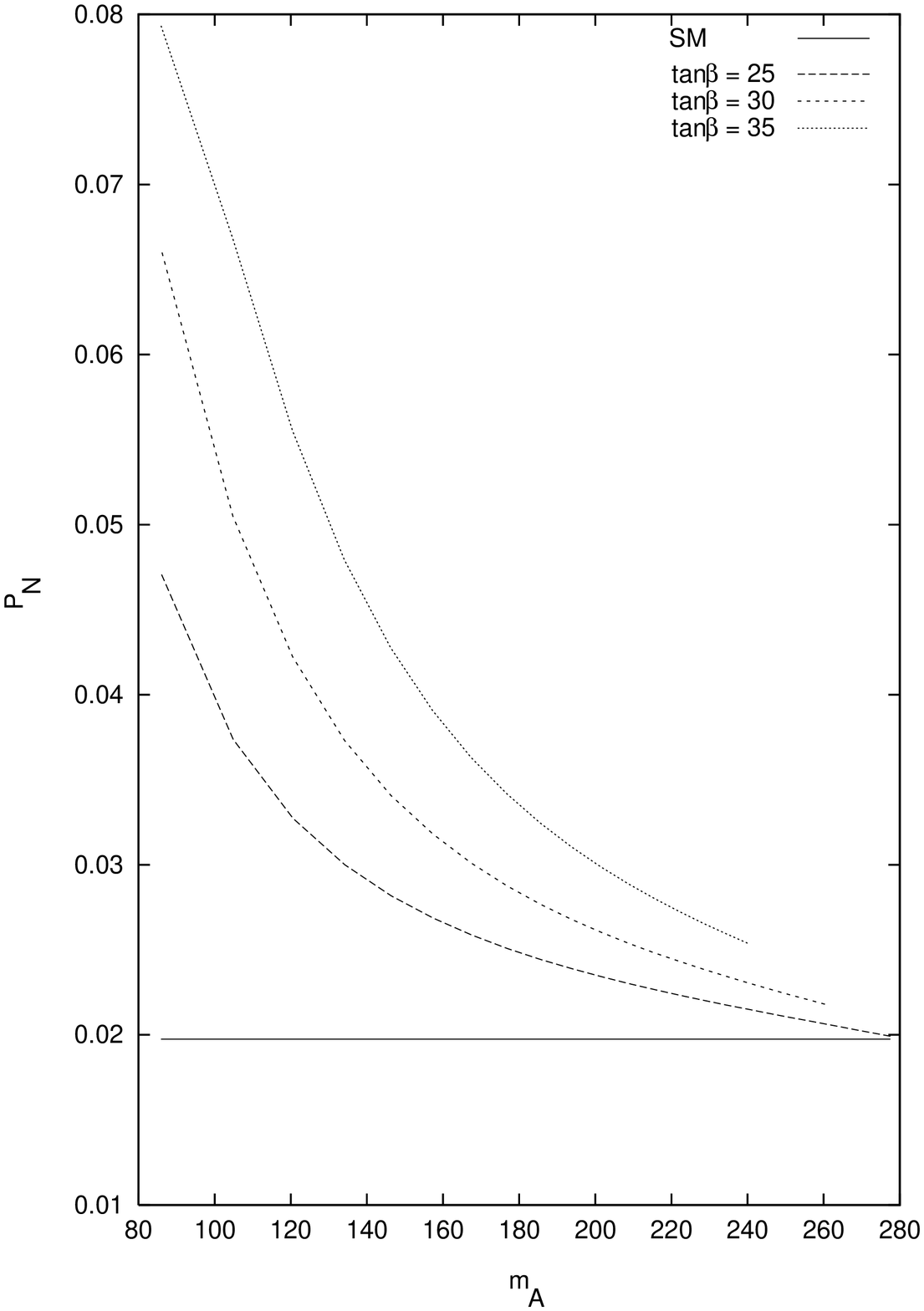,width=6in,height=7in}
\vskip 1cm
\caption{Normal Polarisation asymmetry with $m_A$ in relaxed SUGRA
model. Other parameters are : $m = M = 130 ; A = - 1; \hat{s} =
0.65$.All masses are in GeV}
\label{figure6}
\end{figure}

 Figures \ref{figure1} - \ref{figure6} summarize our results, wherein
we have presented the three polarization values in the SM,minimal
SUGRA and the ``relaxed'' SUGRA (RSUGRA) as discussed before. The extra
parameter in RSUGRA has been taken to be the CP-odd Higgs boson mass $m_A$
which is related to the parameters in the potential by 
\beq
\label{newseven}
m_A^2 = 2 \Delta_0^2 + 2 \mu^2
\eeq
with  the parameters being evaluated at $M_W$. The general comment
about all the three polarizations is that in SUGRA, there is no
appreciable change from the SM value even with NHB contributions. This 
is because at high \tanbeta, the constraints obtained through $b \to s 
\gamma$ limits, forces the three neutral Higgs boson to large mass
value thus suppressing the NHB contributions. This is precisely the
reason that in relaxed version of SUGRA, where low Higgs mass become
allowed, considerable deviations from SM values are possible.
\par Turning now to the absolute values of $P_L,P_T$ and $P_N$ as
shown in Figs \ref{figure1} - \ref{figure6}, it is important to note
that at and around the resonant peaks, the dominant contributions come 
from the resonant B-W contributions, eqn(\ref{two}) multiplied by a
phenomenologically empirical factor $\kappa_v = 2.35$. We have taken
this factor to be universal for all resonances whereas the actual
number is fitted only to $J/\Psi$ production. This introduces some
uncertainty in values of the cross-section around the higher resonances
and it is for this reason that the polarisation values given Figs
\ref{figure1} - \ref{figure6} are more reliable in between the
$c\bar{c}$ resonances rather than at the resonances. Typically for
$tan\beta = 30$ in the region $ 0.63 \leq \hat{s} \leq 0.68$, as well
as $0.77 \leq \hat{s} \leq 0.82$ the longitudinal polarization
increases in magnitude by about $50\%$. A similar pattern occurs for
$P_T$ in the same region and in regions between higher resonances. For 
the normal polarisation $P_N$ in the two regions $0.63 \leq \hat{s}
\leq 0.68$ and $0.77 \leq \hat{s} \leq 0.82$ the value changes by a
factor of two. In general in the regions between the resonances there
are changes in the values of polarizations which are sufficiently
large for experimental detection as and when data become available.Figs
\ref{figure2},~\ref{figure4},~\ref{figure6} shows the general
dependence of the polarization parameters on \tanbeta ~and $m_A$.
\par In conclusion our calculations indicate that in MSSM with a large 
\tanbeta~ and low $m_A$ value, the polarization asymmetries in $B \to
X_s \tau^+ \tau^-$ are sensitive to neutral higgs boson exchange
contributions . Similar kind of enhancements were also claimed in
\cite{nardi1} but there the R-parity violating couplings were
responsible for it, but here we are working in model where R-parity is 
a exact symmetry.
The usefulness of polarization measurements in the
context of the Standard Model and beyond
have already been emphasized in literature
\cite{kruger1,hewett1,nardi1}  and our  results 
are expected to be useful in comparing
SUSY-model predictions with experimental results when they become
available.
\par One of the authors (AG) is thankful to CSIR for financial
support.  

\pagebreak

%\bibliographystyle{unsrt}
%%%%%%%%%%%%%%%%%%%%%%%%%%%%%%
%  Bibliography
%%%%%%%%%%%%%%%%%%%%%%%%%%%%%%

%\including figures
% \input{figures.tex}
%%%%%%%%%%%%%%%%%%%%%%%%%%%%%%
%  Figures 
%%%%%%%%%%%%%%%%%%%%%%%%%%%%%%
%%%%%%%%%%%%%%%%%%%%%%%%%%%%%%%%%%%%%%%%%%%%
%%%%%%%%%%%%%%%%%%%%%%%%%%%%%%%%%%%%%%%%%%%%
%%%%%%%%%%%%%%%%%%%%%%%%%%%%%%%%%%%%%%%%%%%
%%%%%%%%%%%%%%%%%%%%%%%%%%%%%%%%%%%%%%%%%%%
%%%%%%%%%%%%%%%%%%%%%%%%%%%%%%%%%%%%%%%%%%%
%%%%%%%%%%%%%%%%%%%%%%%%%%%%%%%%%%%%%%%%%%%

\end{document}